\documentclass[mathleft
]{an}
\usepackage{graphicx}
\usepackage{times}
\usepackage[latin9]{inputenc}
\usepackage[T1]{fontenc}
\usepackage{lipsum}
\usepackage{multirow}
\begin{document}

% The following seven commands are intended for editorial usage and should be ignored by
% the author(s).
\Pagespan{789}{}% Document's page range. 
% If second parameter is left empty, the last page is computed automatically.
\Yearpublication{2013}%
\Yearsubmission{2012}%
\Month{11}%   
\Volume{999}%  
\Issue{88}% 
% \DOI{This.is/not.aDOI}% 

\title{Doppler imaging of the double-lined active binary V824 Ara}

\author{L. Kriskovics\inst{1}\fnmsep\thanks{Corresponding author:
  \email{kriskovics.levente@csfk.mta.hu}\newline}
%Example 
%for footnote, note the usage of the \texttt{fnmsep}
%command as separator between institute number and footnote mark} 
\and K. Vida\inst{1}
\and Zs. K\H{o}v\'ari\inst{1}
%\and K. Ol\'ah\inst{1}
\and D. Garcia-Alvarez\inst{2,3,4}%\inst{3}\inst{4}
\and K. Ol\'ah\inst{1}
}
\titlerunning{Doppler imaging of the double-lined active binary V824 Ara}
\authorrunning{L. Kriskovics et al.}
\institute{\inst{1}Research Centre for Astronomy and Earth Sciences, Hungarian Academy of Sciences\\
	   \inst{2}Instituto de Astrof\'isica de Canarias, E-38205 La Laguna, Tenerife, Spain\\
	   \inst{3}Dpto. de Astrof\'isica, Universidad de La Laguna, 38206 La Laguna, Tenerife, Spain\\
           \inst{4}Grantecan CALP, 38712 Bre$\mathrm{\tilde{n}}$a Baja, La Palma, Spain
}
\received{30 May 2005}
\accepted{11 Nov 2005}
\publonline{later}


\abstract{
We introduce an iterative spectral disentangling technique combined with Doppler imaging in order to recover surface temperature maps for both components of 
double-lined active binary systems. Our method provides an opportunity to separate spectra of the active components while minimizing the unwanted disturbances
on the given line profile from the other component. The efficiency of the method is demonstrated on real data of the double-lined RS CVn-type binary V824 Ara.
The resulting Doppler images reveal cool spots on the polar regions as well as low-latitude features on both of the stars. Moreover, both components have
hot spots, that are facing each other. This may indicate interconnection between the stellar magnetic fields.
}

\maketitle

\section{Introduction}
\label{sect_intr}
Doppler imaging of components in a double-lined active binary is a challenging task. Disentangling the spectra of the components is essential,
however, fairly difficult, for the very reason of the line profile distortions due to starspots on both components.
Still, there are different strategies to follow. The simplest approach to handle this issue is omitting the blended phases from the inversion process (Strassmeier \& Rice \cite{strassmeier_v824ara}).
But this way sufficiently good phase coverage can easily be lost, together with all the information on the blended phases, where spectra from the two components overlap.

Another possibility is to apply simultaneous Doppler imaging for the two components, however, the number of the inversion parameters becomes nearly double, consequently, the reliability of the results is reduced. But again, in blended phases the separation of distortions originating from the spots on different components is still unsolved (as an attempt, in blended phases usually homogeneous surface is assumed for the secondary component). Among the very few examples in the literature, see e.g., the application of a two-line Doppler imaging code for $\sigma^2$\,CrB by Strassmeier \& Rice (\cite{strassmeier_sigma2crb}), and the Doppler tomography of ER Vul and TY Pyx in Piskunov (\cite{piskunov}). 
Among double-lined active binaries V824\,Ara (HD\,155555) is the most thoroughly studied system.
In the paper by Hatzes \& K\"urster (\cite{hatzes}) rotationally broadened synthetic spectra were fitted
to the observations and the blended line profiles were subtracted before applying Doppler imaging. Their maps showed polar active
regions on both components, and several low latitudinal features. In Strassmeier \& Rice (\cite{strassmeier_v824ara})
the blended phases were simply excluded from the inversion process. Their results revealed also polar spots, but at lower latitudes
spot distribution was different compared to the ones of Hatzes \& K\"urster (\cite{hatzes}). Zeeman--Doppler imaging was applied for V824 Ara by Dunstone el al. (\cite{dun2008}) where
Stokes \emph{I} maps showed polar features as well.
  \begin{table}[t]
\begin{center}
\begin{tabular}{ccccc}
\hline\hline
HJD	&	phase	&	dd-mm-yyyy	& $\Delta\lambda$ [\AA] &	S/N\\
\hline\hline
2450536.611	&	0.55	& 19-09-2002	&	6403--6447	&	121 	\\
2450537.645	&	0.16	& 20-09-2002	&	6403--6447	&	113	\\
2450538.501	&	0.67	& 20-09-2002	&	6403--6447	&	111	\\
2450538.612	&	0.74	& 21-09-2002	&	6403--6447	&	124	\\
2450539.479	&	0.26	& 21-09-2002	&	6403--6447	&	109	\\
2450539.580	&	0.32	& 21-09-2002	&	6403--6447	&	104	\\
2450539.640	&	0.35	& 21-09-2002	&	6403--6447	&	101	\\
2450540.477	&	0.85	& 21-09-2002	&	6403--6447	&	110	\\
2450540.531	&	0.88	& 22-09-2002	&	6403--6447	&	106	\\
2450540.614	&	0.93	& 22-09-2002	&	6403--6447	&	117	\\
%\hline
%536.502343	&	0.48	& 18-09-2002	&	6536-6580	\\
%536.513738	&	0.49	& 19-09-2002	&	6536-6580	\\
%536.629090	&	0.56	& 19-09-2002	&	6536-6580	\\
%538.547417	&	0.70	& 21-09-2002	&	6536-6580	\\
%538.595123	&	0.73	& 21-09-2002	&	6536-6580	\\
%539.516566	&	0.28	& 22-09-2002	&	6536-6580	\\
%539.624699	&	0.34	& 22-09-2002	&	6536-6580	\\
%540.574059	&	0.91	& 23-09-2002	&	6536-6580	\\

\hline
\end{tabular}
\end{center}
\caption{The table gives the HJDs of the observations, the corresponding phases calculated using Eq.~\ref{EQ} from Sect.~\ref{sect_data},
the observing dates, the wavelength regions ($\Delta\lambda$) and the measured signal-to-noise (S/N) ratios.}
\label{obstable}
\end{table}

In this paper we introduce an iterative spectral disentangling technique combined with Doppler imaging in order to recover surface temperature maps for both
components of V824 Ara. In our new approach profiles in the blended phases are treated to preserve information on the surface spot distribution.

%In section \ref{sect_data}, our spectroscopic data is presented. Section \ref{sect_pars} contains the used astrophysical parameters. The method itself is explained in section \ref{sect_method}, and its application on V824\,Ara is presented in \ref{sect_di}. Finally, the results are summarized in section \ref{sect_sum}. 

\begin{figure*}
\centering
\includegraphics[height=45mm]{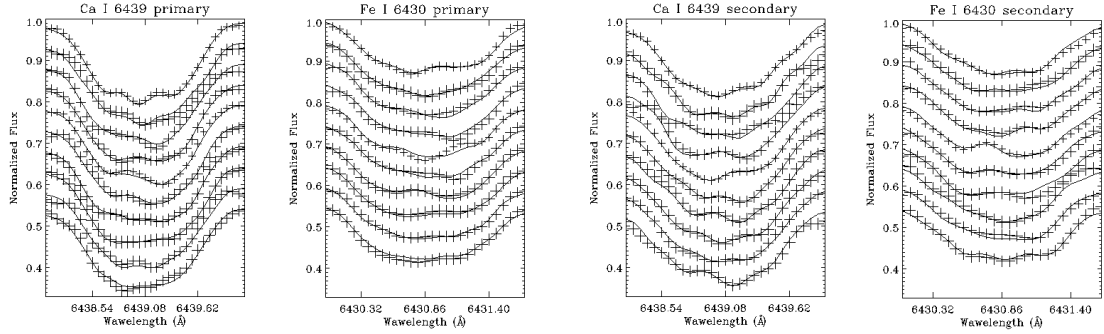}
\caption{Line profile fits (solid lines) to the observations (crosses). From left to right: Ca\,{\sc i} 6439\,\AA\ (1st panel) and Fe\,{\sc i} 6430\,\AA\ profiles (2nd panel) of the primary component, Ca\,{\sc i} 6439\,\AA\ (3rd panel) and Fe\,{\sc i} 6430\,\AA\ profiles (4th panel) of the secondary component. The size of the crosses indicates the S/N.}
\label{spec1}
\end{figure*}

\begin{figure}
\centering
\includegraphics[width=55mm]{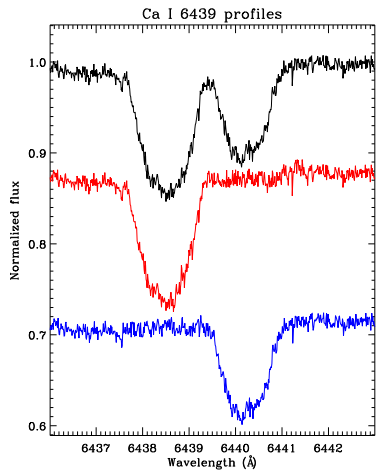}
\caption{An example of disentangled spectra at phase $\phi$=0.411: the original blended spectra from the Ca\,{\sc i} 6439\,\AA\ region (top), and the contributions of the primary (middle) and the secondary (bottom) components. Note that the downward shifts of the contribution spectra are optional.}
\label{spec2}
\end{figure}
\section{The method}
In case of a double-lined binary with two spotted components, the separation of the profiles is troublesome, but essential.
First we set the relative continuum contributions of the two components (which are $0.7$ and $0.3$ for the primary and the secondary components of V824\,Ara, respectively, see Strassmeier \& Rice \cite{strassmeier_v824ara}).
Then, we apply an iterative method, which ensures an effective subtraction even at blended phases where the spectral lines
from the primary and the secondary stars heavily overlap. Our method, step by step:
\begin{itemize}
    \item an initial line profile of the secondary component is calculated by synthesizing an undisturbed line profile (i.e., assuming a homogeneous surface);
    \item the initial profile is subtracted from the composite spectrum providing a preliminary line profile of the primary component;
    \item the set of preliminary profiles are used to perform Doppler imaging (DI) of the primary component;
    \item from the resulting image line profiles of the primary component are calculated with using the forward version of our DI code;
    \item profiles from the forward DI process are subtracted from the original spectra to get profiles of the secondary component;
    \item resulting profiles are used to perform Doppler imaging for the secondary component;
    \item Doppler maps of the secondary star are used as input to calculate distorted line profiles of the secondary with the forward DI code;
    \item these profiles of the secondary are subtracted from the original spectra;
    \item resulting spectra are used to derive Doppler maps for the primary component.
\end{itemize}
The above steps are repeated iteratively until the Doppler images are converged.

\begin{table}
\begin{center}
\begin{tabular}{lll}
\hline\hline
Parameter & Primary & Secondary\\
\hline\hline
$T_{\mathrm{eff}}$    &    $5400\pm100\,\mathrm{K}$ & $5040\pm150\,\mathrm{K}$\\
$\log{g}$        &    $4.0\pm0.5$ & $4.5\pm0.5$ \\
$v\sin{i}$        &    $36.9\pm1.0\,\mathrm{km/s}$ & $33.5\pm1.0\,\mathrm{km/s}$\\
Inclination  $i$      &   $55^{\circ}\pm5$     &   $55^{\circ}\pm5$         \\
Microturbulence $\zeta_{Ca}$ &    $1.5\,\mathrm{km/s}$ & $1.5\,\mathrm{km/s}$    \\
Microturbulence $\zeta_{Fe}$ &    $2.0\,\mathrm{km/s}$ & $1.5\, \mathrm{km/s}$    \\
\hline
\end{tabular}
\end{center}
\caption{Basic astrophysical parameters for V824\,Ara adopted for Doppler reconstructions from Strassmeier \& Rice (\cite{strassmeier_v824ara}).}
\label{orb}
\end{table}
\begin{figure*}
\centering
\includegraphics[width=150mm]{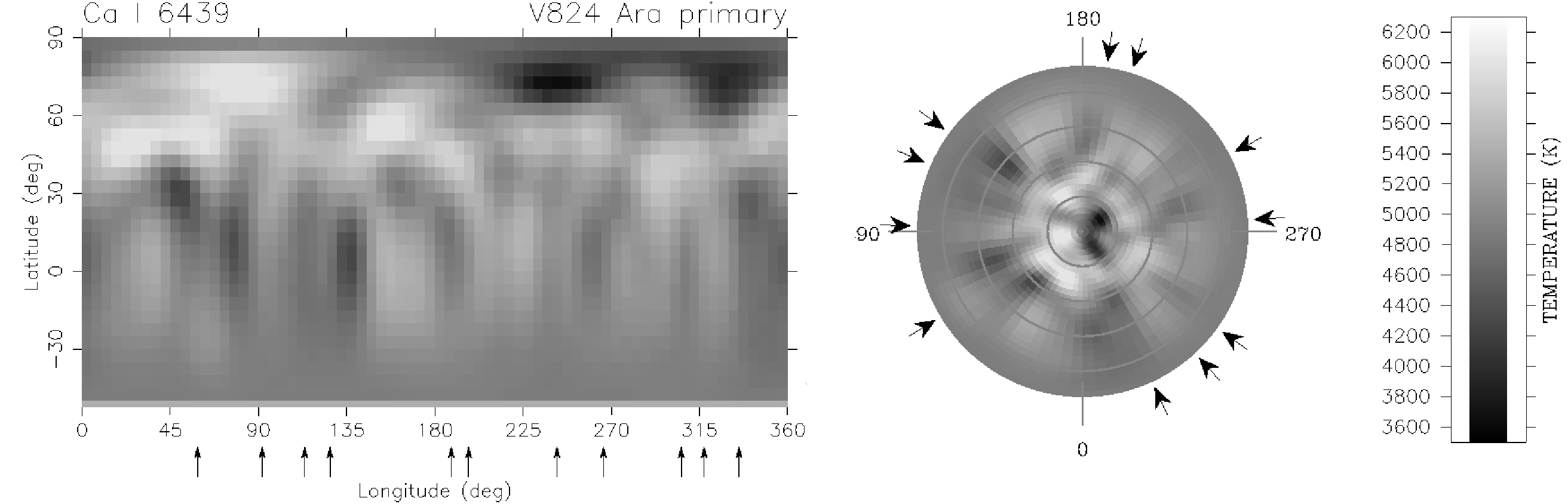}
\caption{Resulting Doppler reconstruction of the primary component of V824\,Ara for the Ca\,{\sc i} 6439\,\AA\ line.
The surface temperature map is plotted in pseudo-Mercator projection (left) and in pole-on view (right).
Arrows mark the phases of the observations.}
\label{di1ca}
\end{figure*}

\begin{figure*}[]
\centering
\includegraphics[width=150mm]{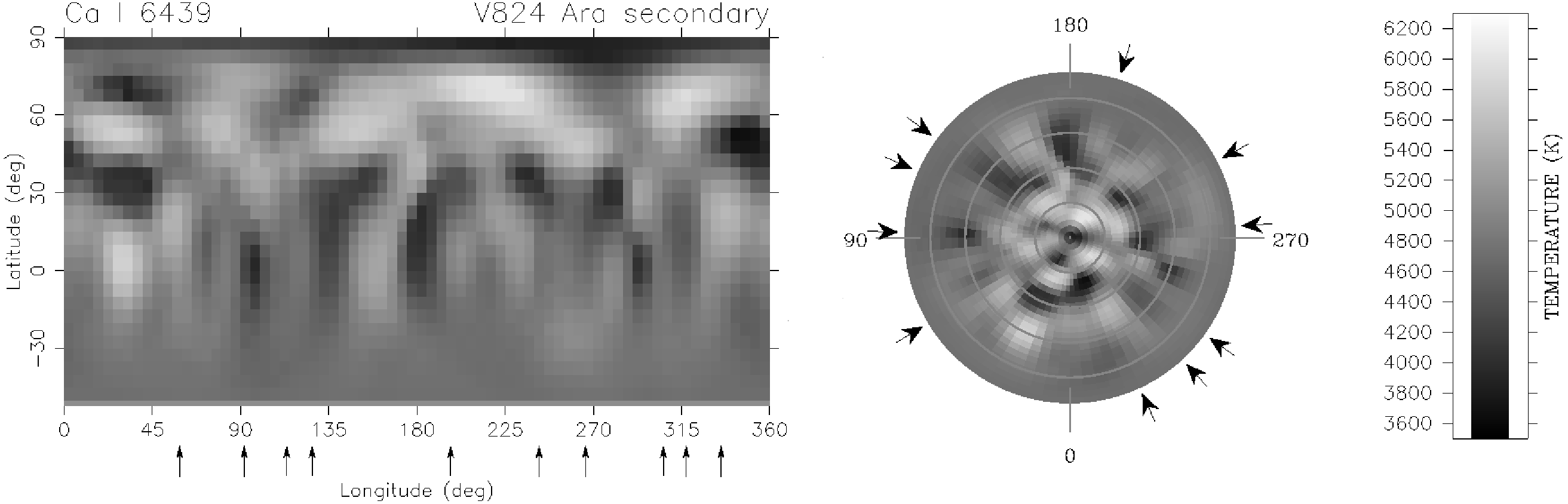}
\caption{Resulting Doppler reconstruction of the primary component of V824\,Ara for the Ca\,{\sc i} 6439\,\AA\ line.
The surface temperature map is plotted in pseudo-Mercator projection (left) and in pole-on view (right).
Arrows mark the phases of the observations.}
\label{di2ca}
\end{figure*}

\section{Doppler imaging of V824\,Ara}
\label{sect_data}
\subsection{Spectroscopic data}
10 spectra were taken between 19 and 22 Sept 2002 with the Coud\'e-Echelle Spectrometer and with the Loral $2688\times512$ $15\mu$ CCD mounted on the ESO $3.6\,\mathrm{m}$ telescope
(3P6) in La Silla, Chile. The observations were carried out in single-order mode covering the $6400-6450$\,\AA range with the exposure
time of 900\,s, typically providing signal-to-noise ratios of 100--125 at peak resolution $R=65,000$. The gathered data cover one orbital
cycle with suitable phase sampling to perform Doppler imaging. Table \ref{obstable} summarizes the observing log. Phases are calculated using the following equation (Strassmeier \& Rice \cite{strassmeier_v824ara}):
\begin{equation}\label{EQ}
\rm{HJD}= 2,446,998.4102 + 1.6816 \times E. \\
\end{equation}
Data reduction was carried out using the standard NOAO/IRAF
\footnote[1]{IRAF is distributed by the National Opctical Astronomy Observatory, which is operated by the Association of Universities
for Research in Astronomy (AURA) under cooperative agreement with the National Science Foundation.
}
routines.

\subsection{Astrophysical parameters}
\label{sect_pars}

Basic input parameters for Doppler imaging were adopted from Strassmeier \& Rice (\cite{strassmeier_v824ara}).
However, the inclination $i$ and projected equatorial velocity $v\sin{i}$ values are fine-tuned
with a grid search method based on finding the best fit $\chi^2$ from Doppler reconstructions over
a reasonable part of the parameter plane (for the method see Unruh \cite{unruh}).
Accordingly, for this study we used $v\sin{i}=36.9\,\mathrm{km/s}$ for the primary, and $v\sin{i}=33.5\,\mathrm{km/s}$ for the secondary component. The former values is slightly higher, while the latter is lower compared to the ones in Strassmeier \& Rice (\cite{strassmeier_v824ara}). The errors of the derived $v\sin{i}$ and $i$ are $\pm 1\,\mathrm{km/s}$ and $\pm 5^{\circ}$, respectively. These values are between the previous error estimates given by Strassmeier \& Rice (\cite{strassmeier_v824ara}).

\begin{figure*}
\centering
\includegraphics[width=150mm]{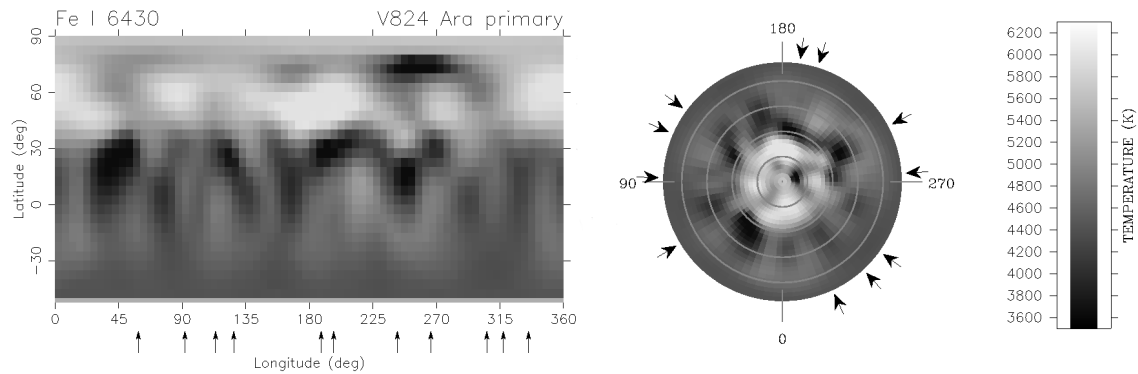}
\caption{Resulting Doppler reconstruction of the primary component of V824\,Ara for the Fe\,{\sc i} 6430\,\AA\ line.
The surface temperature map is plotted in pseudo-Mercator projection (left) and in pole-on view (right).
Arrows mark the phases of the observations.}
\label{di1fe}
\end{figure*}

\begin{figure*}[]
\centering
\includegraphics[width=150mm]{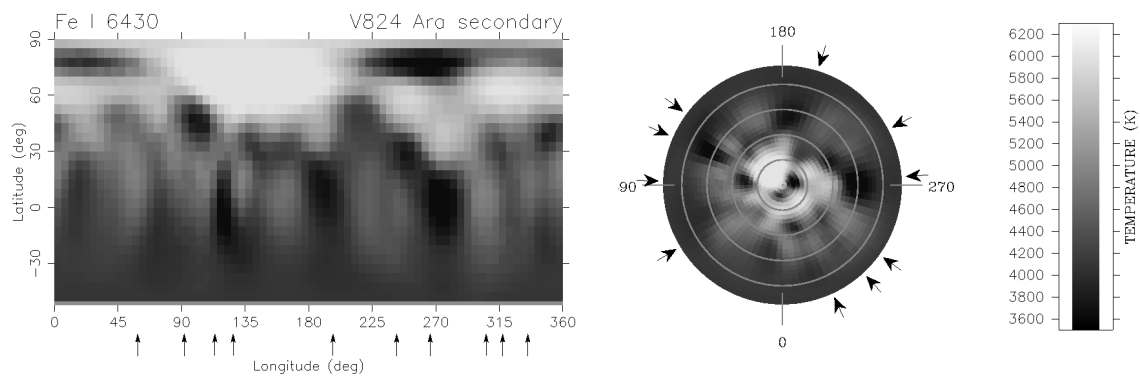}
\caption{Resulting Doppler reconstruction of the secondary component of V824\,Ara for the Fe\,{\sc i} 6430\,\AA\ line.
The surface temperature map is plotted in pseudo-Mercator projection (left) and in pole-on view (right).
Arrows mark the phases of the observations.}

\label{di2fe}
\end{figure*}
\subsection{Doppler images of V824 Ara}\label{sect_di}

Doppler surface reconstructions are carried out with the $\texttt{TempMap}$ code originally developed by Rice et al. (1989).
The program performs a full LTE spectrum synthesis by solving the equation of transfer through a grid of ATLAS-9 (Kurucz \cite{kurucz}) model atmospheres at all aspect angles and for a given set of chemical abundances.
For the synthetic line profiles, the oscillator strengths ($\log{gf}$) and excitation potentials were taken from the VALD database (Piskunov et al. \cite{vald1}, Kupka et al. \cite{vald2}).

For the mapping we select the most commonly used absorption lines with well-known formation physics and mostly free of telluric blends in the vicinity: the Ca\,{\sc i} 6439\,\AA\ and Fe\,{\sc i} 6430\,\AA\ lines. We note that in some of the orbital phases at Fe\,{\sc i} 6430\,\AA\ of the secondary star there is still a severe blending of the Fe\,{\sc ii} 6432\,\AA\ line which can be eliminated in the iterative process.
Resulting Doppler maps for the Ca line are plotted in Fig.~\ref{di1ca} (primary comp.) and Fig.~\ref{di2ca} (secondary comp.), while Fe line reconstructions are shown in Fig.~\ref{di1fe} (primary) and Fig.~\ref{di2fe} (secondary).
The Ca and Fe line reconstructions reveal similar surface temperature distributions, i.e., mostly cool spots near the poles on both components (more significantly on the the primary star), and sporadic cool spots at lower latitudes. Bright features are also present and are somewhat stronger on the primary component.

\section{Summary and discussion}
\label{sect_sum}
Surface Doppler reconstruction of the components in such a doubly active binary system like our target V824\,Ara is still a difficult task
because of the overlapping spectra and the larger (twofold) number of astrophysical and atomic parameters. Despite that,
we successfully carried out an iterative method which solved the separation of the composite spectra,
even when they are disturbed by surface spots. At the end, our step by step method resulted in reliable Doppler reconstructions for both
components.
The revealed cool spots near the visible poles of both components are in agreement with the previous findings (Hatzes \& K\"urster 1999, Strassmeier \& Rice \cite{strassmeier_v824ara}). On the other hand, low-latitude features seem to be more variable on a longer timescale, as compared to the maps from 1990 (Hatzes \& K\"urster \cite{hatzes}) and from 1996 (Strassmeier \& Rice \cite{strassmeier_v824ara}).
Our reconstructions show that bright features appear on the opposing hemispheres. Since these features can be seen on both the Ca\,{\sc i}~6439\,\AA\ and Fe\,{\sc i}~6430\,\AA\ maps, we believe that these are indeed real. Strong interactions between the magnetic fields of the members of close RS CVn-type binaries are usual (cf., e.g., Uchida \& Sakurai \cite{uchida}, Siarkowski \cite{siarkowski}). We think that the hot spots in our maps (Figs.~\ref{di1ca}--\ref{di2fe}) on the components facing each other may indicate such interactions.

\small{
\emph{
Acknowledgements.} LK, KV, ZsK and OK are grateful to the Hungarian Science Research Program (OTKA) for support under the grant K-81421. This work is supported by the "Lend\"ulet-2009" and "Lend\"ulet-2012" Young Researchers' Programs of the Hungarian Academy of Sciences and by the HUMAN MB08C 81013 grant of the MAG Zrt. 
}

\end{document}